\documentclass[a4paper,12pt]{article}
\usepackage[utf8]{inputenc}
\usepackage[T2A]{fontenc}
\usepackage[english]{babel}
\usepackage{geometry}
\usepackage{fancyhdr}
\usepackage{amsmath,amsfonts,amsthm,amssymb}
\usepackage{graphicx} 
\usepackage{hyperref}
\usepackage{amssymb,graphicx,float,amsmath,amsfonts,amsthm,cases}
\usepackage{mathtools}
\usepackage[symbol*]{footmisc}
\graphicspath{{/}}
\usepackage{tensor}
\usepackage[numbers]{natbib}
\usepackage{float}
\usepackage{subfig}

\begin{document}
\vspace{10pt}

\begin{center}
	{\LARGE \bf A survey of ultra-compact rotating boson star spacetimes}
	
	\vspace{20pt}
	
	N. Sukhov$^{a,b}$\footnote[1]{\textbf{e-mail:} nsukhov@princeton.edu},
%	F. Pretorius$^{a,b}$\footnote[2]{\textbf{e-mail:} fpretori@princeton.edu}
	\vspace{15pt}
	
	$^a$\textit{Department of Physics, Princeton University, Princeton, New Jersey, 08544, USA}\\
	$^b$\textit{Princeton Gravity Initiative, Princeton University, Princeton, New Jersey, 08544, USA}\\	
	\vspace{5pt}
	
\end{center}

\vspace{5pt}

\begin{abstract}
	Solitonic boson stars (SBS) are compact shell-like objects with an inside having a nearly constant value of scalar field bounded by a thin shell where the scalar field rapidly changes. While the spherically symmetric SBS can be described by an analytical approximation that works well in the thin-shell case, no such approximation exists for rotating SBS and the investigation of such stars poses a numerical challenge. We numerically investigate rotating SBS in case if the relative thickness of the shell is small. We compute Gerosh-Hansen multipole moments to study the structure of SBS spacetime in the outside region and the closeness of SBS to the rotating black hole metric.
\end{abstract}

\section{Introduction}
One of the most remarkable recent advances in astrophysics is the detection of gravitational waves (GW) produced by binary Black Hole (BH) and Neutron Star (NS) systems \citep{LIGOScientific:2016aoc, LIGOScientific:2017vwq, LIGOScientific:2019fpa, LIGOScientific:2020ibl, LIGOScientific:2021qlt}. This discovery has started an entirely new area of astrophysics often called gravitational wave astronomy. One of the fundamental challenges of this emerging area is the exploration of other yet unobserved classes of objects that are capable of producing detectable GW. Such objects must be compact and have strong gravity, described by Einstein's General Theory of Relativity (GR) and are typically called exotic compact objects or ECO \citep{Giudice:2016zpa}.

We explore one of the most established classes of such objects called Boson Stars (BS). The BS are described by mathematically self-consistent and well-posed equations within GR, and the astrophysical processes that include BS, such as star mergers or the gravitational collapse, can be simulated numerically by directly solving Einstein's equations starting from generic initial data \citep{Palenzuela:2017kcg, Bezares:2017mzk, Siemonsen:2023hko}. Although, the particles, massive scalar bosons, that could form a BS have not been experimentally observed, they fall within the well-understood framework of particle physics and could potentially be included in an extension of the Standard Model of particle physics and constitute a fraction of the dark matter (DM) \citep{Bertone:2019irm}. There is also a great variety of compact BS models some of which have unusual features such as peculiar optical properties \citep{Grandclement:2016eng} or GW waveform profile emissions in binary mergers \citep{Siemonsen:2020hcg} that could make them distinguishable from BH or NS \citep{Cardoso:2016oxy}.

The most common kind of BS consists of a single complex scalar field with some scalar field potential $V(\Phi)$, there have been numerous potentials proposed in the literature \citep{Kaup:1968zz, Colpi:1986ye, Friedberg:1986tq, Kleihaus:2005me, Kleihaus:2007vk, Schunck:1999zu, Balakrishna:1997ej}. In this article we focus on the solitonic boson stars \citep{Friedberg:1986tq}(SBS). The SBS models have the scalar field potential that has at least two vacua, the star in such case has a shell-like structure; the inside and the outside regions of the star have nearly uniform scalar field absolute values corresponding to different vacua and are separated by a thin shell where the scalar field rapidly transitions from one vacuum to another. The star itself is prevented from gravitational collapse by the surface tension in the shell \citep{Boskovic:2021nfs}. The simplest kind of solitonic potential that we are going to cover in the present article is given by the equation \eqref{Description:solitonic_potential}
\begin{equation}
	\label{Description:solitonic_potential}
	V(|\Phi|^2) = \mu^2 |\Phi|^2 \left(1 - \frac{2|\Phi|^2}{\sigma_0^2}\right)^2.
\end{equation}

The spherically symmetric SBS have a particularly simple structure that can be described with an analytical approximation \citep{Friedberg:1986tq}. The relative thickness of the boundary shell is controlled by the stiffness of the potential set by the parameter $\sigma_0$, the limit $\sigma_0 \to \infty$ corresponds to the case of an infinitely thin shell. However, the rotating SBS have a much more complicated boundary shell that does not admit a simple analytical approximation. In this article we investigate the thin-shell limit of rotating SBS. To obtain such solutions we have developed a novel numerical pseudo-spectral method that splits the computational domain into a series of patches to provide sufficient resolution to the boundary region. The details of the method are provided in the companion article \citep{Sukhov:bstars}. We obtain several families of SBS in a wide range of the parameter $\sigma_0$ and investigate the basic properties of the stars, such as mass, compactness, charge and angular momentum and the presence of ergoregions. We use the usual definition of compactness $\mathcal{C} = M/R$, where $M$ and $R$ are the object mass and radius, the compactness of a non-rotating Schwarzschild BH is $\mathcal{C} = 1/2$.

Deviations of the GW inspiral signals from that of a BH can be traced to finite-size effects, which reflect the inner properties of the object. The leading order post-Newtonian inner structure effects can be described in terms of the quadrupole moment $M_2$ \citep{Poisson_Will_2014}. In GR, the multipole moments (or Gerosh-Hansen (GH) moments or indices) \citep{Geroch:1970cd, Gerosh:1971, Hansen:1974zz} generalize the multipole moments of the Newtonian gravity. They can be used to describe the spacetime around a compact object and therefore encode the properties of the physical processes that happen in its vicinity, for instance, the GW emissions from a small body falling on a larger object \citep{Ryan:1995wh}. 

There exist two sets of multipole moments; one characterizes the mass distribution of an object and the other describes its angular momentum (current) distribution. The multipolar structure of an axisymmetric spacetime can be described by a series of scalars $M_n$ and $S_n$ corresponding to the mass and current moments respectively. The multipole moments of the Kerr spacetime take an especially simple form
\begin{equation}
	\label{Description:Kerr_moments}
	M_{2n} = (-1)^{n + 1} M a^{2n},\qquad M_{2n+1} = 0,\qquad S_{2n} = 0,\qquad S_{2n+1} = (-1)^n M a^{2n + 1},
\end{equation}
where $M$ is the black hole mass and $a = J/M$ is the spin parameter. Any deviation of the multipolar structure signals that the spacetime is not described by the Kerr metric, therefore the measurement of the GH moments can be used as a null-hypothesis test of the Kerr metric \citep{Ryan:1995wh, Barack:2018yly, Cardoso:2019rvt, Gair:2012nm, Yunes:2013dva, Berti:2015itd, Cardoso:2016ryw}.

The GH moments have been considered for rotating BS spacetimes in \citep{Ryan:1996nk, Vaglio:2022flq, Adam:2022nlq, Adam:2023qxj}, however no analysis was done for the most compact and therefore potentially closest to Kerr SBS. In this article we investigate the multipolar structure of the rotating SBS and study how close the metric outside the star resembles Kerr metric. We then explore what happens in the thin-shell limit.

This article is structured as following. In the next section we outline the theory behind the solitonic boson stars and their properties. In Section 3 we give a brief exposition of the numerical methods we use to obtain SBS spacetimes, a more detailed discussion of the methods can be found in the companion paper \citep{Sukhov:bstars}. In Section 4 we explore SBS families in the wide range of $\sigma_0$ parameter and comment on the thin-shell limit $\sigma_0\to 0$. We conclude in section 5.

\section{Model}
We consider a system of Einstein-Klein-Gordon equations, where a complex scalar field is minimally coupled to gravity. The system is described by the action
\begin{equation}
	\label{Model:Action}
	S = \int d^4x \left[\frac{R}{2\kappa}- g^{ab} \partial_a \Phi^* \partial_b \Phi - V(|\Phi|^2)\right],
\end{equation}
where $\Phi$ is the complex scalar field with the potential $V(|\Phi|^2)$ given by \eqref{Description:solitonic_potential}, $g$ is the metric determinant and $R$ is the Ricci scalar. Varying the action with respect to the field and the metric we obtain the system of field equations
\begin{subequations}
\begin{align}
	\label{Model:Einstein_equations}
	R_{ab} - \frac{1}{2} R g_{ab} = \kappa T_{ab},\\
	\label{Model:KG_equations}
	\square\Phi - V'(|\Phi|^2)\Phi = 0,
\end{align}
\end{subequations}
where $T_{ab}$ is the canonical stress-energy tensor
\begin{equation}
	\label{Model:Stress-Energy tensor}
	T_{ab} = \partial_a \Phi \partial_b \Phi^* + \partial_a \Phi^* \partial_b \Phi - g_{ab}\left[g^{cd} \partial_c \Phi \partial_d \Phi^* + V(|\Phi|^2)\right].
\end{equation}
Rotating boson stars are axisymmetric time periodic solutions that are described by the ansatz
\begin{equation}
	\label{Model:Scalar_field_ansatz}
	\Phi(t,r,\theta,\phi) = \phi(r,\theta)\,  e^{i \Omega t + i m \varphi},
\end{equation}
where $\Omega \in \mathbb{R}$ is the scalar field oscillation frequency and $m \in \mathbb{N}$ is so-called \textit{azimuthal harmonic number} or \textit{azimuthal winding number}. We take the following widely used stationary axisymmetric ansatz for the metric
\begin{equation}
	\label{Model:KEH_ansatz}
	ds^2 = - e^{\gamma + \rho}dt^2 + e^{2\sigma}(dr^2 + r^2 d\theta^2) + e^{\gamma - \rho} r^2 \sin^2\theta (d\varphi - \omega dt)^2.
\end{equation}
The set of explicit differential equations for the metric functions and the scalar field together with the appropriate boundary conditions is given in the Appendix. We are free to choose the length scale and the energy scale of the system which allows us to set $\kappa = 1$ and $\mu = 1$. This choice effectively leaves $\sigma_0$ as the sole independent action parameter. \footnote{The other popular choice $\kappa = 8\pi$ will lead to a different normalization for $\sigma_0$, in particular $\left.\sigma_0\right|_{\kappa = 1} = \sqrt{8\pi} \left.\sigma_0\right|_{\kappa = 8\pi}$.}

We now consider the symmetries of the system. The metric \eqref{Model:KEH_ansatz} has two Killing vectors
\begin{equation}
	\xi = \partial_t,\qquad \nu = \partial_\varphi.
\end{equation}
which allows us to use Komar formulae \citep{Wald:1984rg} to define spacetime mass and angular momentum
\begin{subequations}
\label{Model:mass_momentum}
\begin{align}
&M = \frac{\kappa}{4\pi} \int_\Sigma\left(T_{ab} - \frac{1}{2}T g_{ab}\right)n^a \xi^b d V,\\
&J = - \frac{\kappa}{8\pi} \int_\Sigma\left(T_{ab} - \frac{1}{2}T g_{ab}\right)n^a \eta^b dV,
\end{align}
\end{subequations}
where $\Sigma$ is an asymptotically flat timelike hypersurface $t = const$, $n^a$ is the unit normal vector to the hypersurface $n^a n_a = -1$ and $dV$ is the volume element of the hypersurface. We provide the explicit expressions for the mass and the angular momentum in terms of the metric functions and the scalar field in the Appendix.

The action \eqref{Model:Action} is invariant under $U(1)$ transformations
\begin{equation}
	\Phi \to e^{i \alpha} \Phi,
\end{equation}
which leads to the presence of a conserved current
\begin{equation}
	j_a = i (\partial_a \Phi^* \Phi - \Phi^* \partial_a \Phi),\qquad j^a_{;a} = 0,
\end{equation}
where semicolon denotes the covariant differentiation, and a conserved charge
\begin{equation}
	Q = \int_\Sigma j_a n^a d V.
\end{equation}
Following \citep{Schunck:1996he} we note a connection between the stress-energy tensor and the current $n^a T_{a \varphi} = m n^a j_a$, since $\partial_\varphi \Phi = i m \Phi$, this leads to the quantization relation between the angular momentum and the charge
\begin{equation}
J = - \frac{\kappa m}{8\pi} Q.
\end{equation}

To characterize how close the spacetime outside the star is to a black hole solution we compute the Gerosh-Hansen (GH) moments. The first four non-zero GH moments for the metric \eqref{Model:KEH_ansatz} are given by the expressions
\begin{subequations}
	\label{Model:GH_moments}
	\begin{align}
		&M_2 = \frac{1}{2}\rho_2 + \frac{M^3}{3}\left(1 + 4 b_0\right),\\
		&S_3 = - \frac{3}{4}\omega_2 + \frac{3}{5} j M^4 \left(1 + 4 b_0\right),\\
		&M_4 = \frac{1}{2}\rho_4 - \frac{4}{7} \rho_2 M^2 \left(1 + 3 b_0\right) + M^5 \left[\frac{8}{7}b_2 - \frac{16}{5}b_0^2 - \frac{32}{31} b_0 + \frac{6}{35}j^2 - \frac{19}{105}\right],\\
		&S_5 = \frac{5}{6}\omega_4 - \frac{5}{6} \omega_2 M^2 \left(1 + 4 b_0\right) + \frac{5}{21}\rho_2 j M^3 - \\
		&\qquad\qquad j M^6 \left[\frac{40}{21}b_2 - \frac{48}{7}b_0^2 - \frac{208}{63} b_0 + \frac{2}{7}j^2 - \frac{25}{63}\right],
	\end{align}
\end{subequations}
where $b_n = B_n/M^{2n+2}$ and $j = J/M^2$, and $\rho_n$, $\omega_n$ or $B_n$ are asymptotic metric expansion coefficients
\begin{subequations}
	\label{Model:KEH_asymptotics}
	\begin{align}
		&e^\gamma = 1 + \frac{B_0}{r^2} + \sum\limits_{n = 1}^\infty \frac{B_n}{r^{2n + 2}} P^{(\frac{1}{2},\frac{1}{2})}_n(\cos\theta),\\
		&\rho = \sum\limits_{n = 0}^\infty\left[\frac{\rho_n}{r^{2 n + 1}} + O(r^{-(2n + 2)}) \right] P_n(\cos\theta),\\
		&\omega = \sum\limits_{n = 0}^\infty\left[\frac{\omega_n}{r^{2n + 3}} + O(r^{-(2n + 4)}))\right] P^{(1,1)}_n(\cos\theta),
	\end{align}
\end{subequations}
where $P^{(\frac{a}{2},\frac{a}{2})}_n(\cos\theta)$ are Jacobi polynomials. These coefficients can be efficiently computed by numerically taking the limits
\begin{subequations}
	\begin{align}
		&B_n = \frac{2n + 1}{2} \lim\limits_{r\to\infty} r^{n+2}\int\limits_0^\pi \left(e^{\gamma(r,\theta)}-1\right) P^{(\frac{1}{2},\frac{1}{2})}_n(\cos\theta) d\theta,\\
		&\rho_n = \frac{(n + 1)}{2} \frac{n! (n+1)!}{\Gamma\left(n + \frac{3}{2}\right)^2} \lim\limits_{r\to\infty} r^{n+1}\int\limits_0^\pi \rho(r,\theta) P_n(\cos\theta) d\theta,\\
		&\omega_n = \frac{(2n + 3)(n+2)}{8(n+1)} \lim\limits_{r\to\infty} r^{n+3}\int\limits_0^\pi \omega(r,\theta) P^{(1,1)}_n(\cos\theta) d\theta,
	\end{align}
\end{subequations}
where $P_n(\cos\theta) = P^{(0,0)}_n(\cos\theta)$ are Legendre polynomials. We give more exposition on GH moments and the ways to compute it in the Appendix. Finally, it is convenient to normalize the moments
\begin{equation}
	m_{2n} = (-1)^{n+1} \frac{M_{2n}}{M^{2n+1} j^{2n}},\qquad s_{2n+1} = (-1)^n \frac{J_{2n+1}}{M^{2n+2} j^{2s+1}},
\end{equation}
where $M$ is the black hole mass and $a = J/M$ is the spin parameter, so that all of the non-vanishing moments are unit for Kerr. We will call these quantities the reduced GH moments.

\section{Numerics}
In order to solve the BS equations numerically, we employ the method we describe in detail in the companion paper \citep{Sukhov:bstars}. We use the multidomain Chebyshev collocation pseudo-spectral method to obtain a BS solution starting from an initial guess and add necessary resolution to areas where the solution is stiff. As an initial guess for smooth enough solutions we use a seed that is similar to the one described in \citep{Grandclement:2014msa}
\begin{equation}
	\begin{split}
		&\gamma(r,\theta) = \rho(r,\theta) = \sigma(r,\theta) = \omega(r,\theta) = 0,\\
		&\phi(r,\theta) = f_0 \rho^2 e^{-\frac{\rho^2}{s_x^2} - \frac{z^2}{s_z^2}},
	\end{split}
\end{equation}
where
\begin{equation}
	\label{Result:cylindrical_coordinates}
	\rho = r \cos\theta,\quad z = r\sin\theta
\end{equation}
are cylindrical coordinates. An example of a suitable choice of parameters is $s_x = 4$, $s_z = 3$ and $f_0 = 0.03$ for a star with $\sigma_0 = 1$, $m = 1$ and $\Omega = 1$. Knowing several solutions we can extrapolate the star profile to the region with more compact and stiff stars and use the extrapolation result as the next seed. We provide the detailed description of the process in \citep{Sukhov:bstars}. To control the numerical error we compute $L_\infty$ norm of the independent residuals, which means we take the highest overall absolute value of the residuals in the computational domain. We keep the value of the $L_\infty$ at or below $10^{-4}$.

\section{Rotating stars families}
\begin{figure}[H]
	\begin{center}
		\includegraphics[width=1\linewidth]{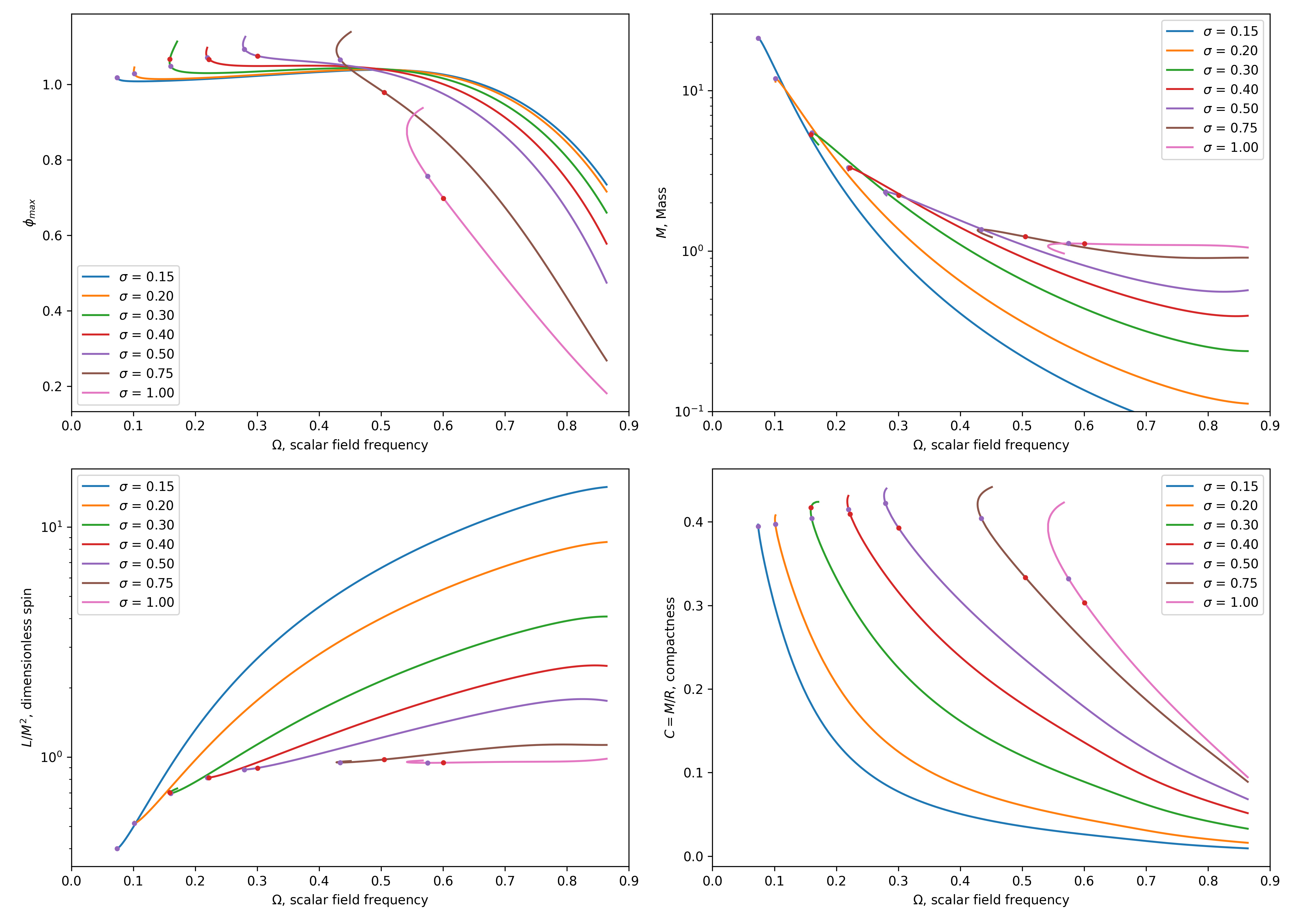}
	\end{center}
	\caption{Rotating SBS diagrams for different values of $\sigma_0$. We have not explored the curves past the mass and the charge maxima ($\Omega < \Omega_{max}$) since those regions of the families are unstable. The red dots correspond to the first occurrences of ergoregions, the purple dots correspond to the mass extrema and thus the ends of the stable parts of BS families.}
	\label{Fig:rotating_star_tracing}
\end{figure}

We restrict our discussion to the case $m = 1$ as the most physically interesting one, although we tested the method for cases $m = \{1, 2, 3, 4\}$. Given the fixed values $m$ and $\sigma_0$ the set of stars forms a single parameter family. We present the $\phi_c$ vs $\Omega$, $M$ vs $\omega$, $J/M^2$ vs $\Omega$ and compactness $\mathcal{C} = M/R$ vs $\Omega$ diagrams for $\sigma_0 = \{0.15, 0.2, 0.3, 0.4, 0.5, 0.75, 1\}$ on the Figure~\ref{Fig:rotating_star_tracing}. The radius $R$ corresponds to the radius containing 99\% of the mass according to the Komar integral. The $\phi_c$ stands for the global maximum of the absolute value of the scalar field.

\begin{figure}[H]
	\begin{center}
		\includegraphics[width=1\linewidth]{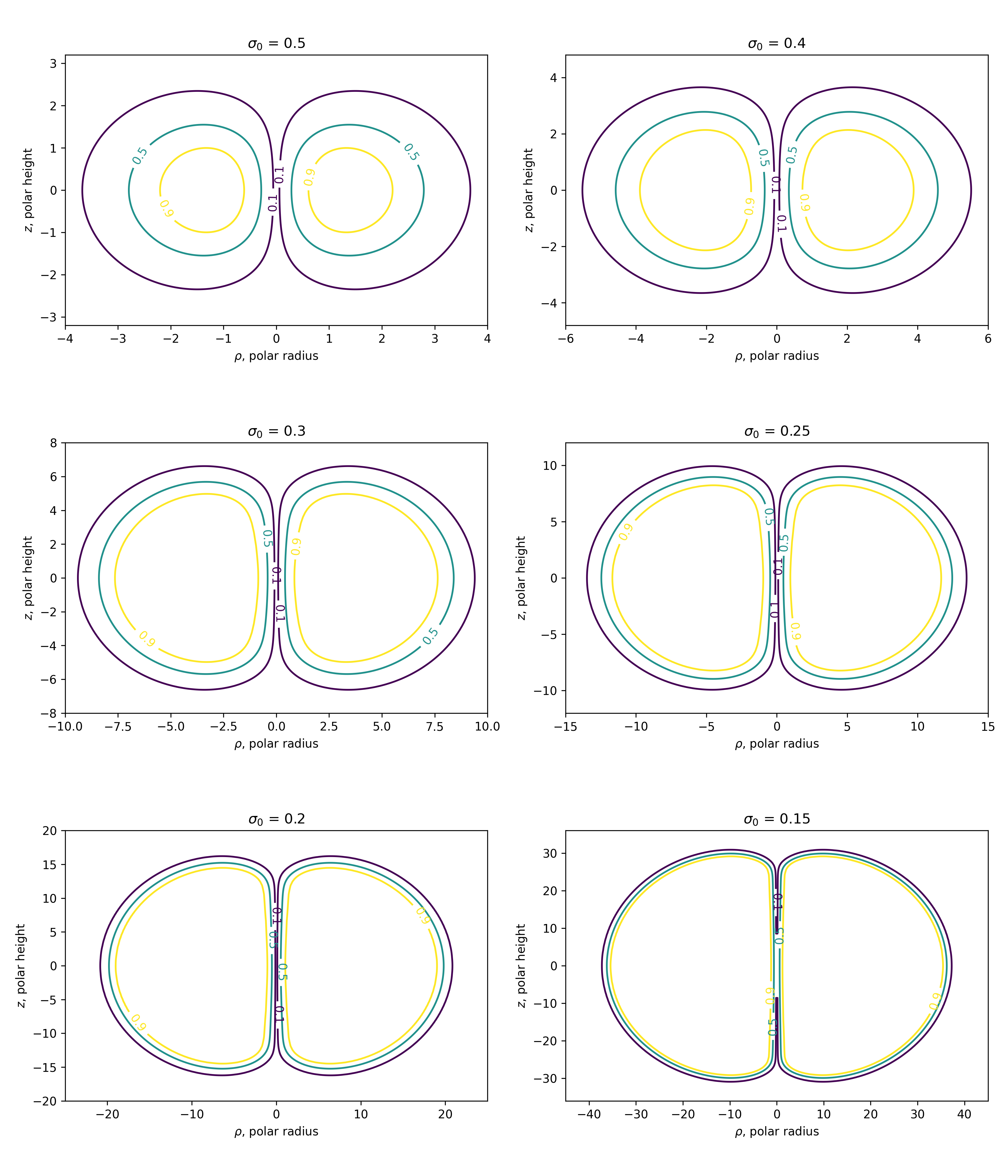}
	\end{center}
	\caption{Rotating SBS scalar field profiles for different values of $\sigma_0$. The stars are selected at the mass extremum points of rotating SBS families. The figures use cylindrical coordinates \eqref{Result:cylindrical_coordinates}. The level lines on the figures indicate the levels of constant absolute value of the scalar field, the values are given in fractions of $\phi_c$.}
	\label{Fig:various_stars}
\end{figure}

According to the analysis \citep{Kleihaus:2011sx} and the numerical results \citep{Siemonsen:2020hcg} we can presume that the point of transition to instability happens on or before the family reaches the global mass maximum (it may happen earlier if the stars develop ergoregions, we will discuss the ergoregions in detail later), we shall call the scalar field frequency $\Omega$ at the global mass maximum point $\Omega_{max}$. According to \citep{Siemonsen:2020hcg} if no ergoregion is present the part of the family immediately to the right of the global mass maximum $\Omega > \Omega_{max}$ is likely to be stable. Therefore as it is evident from Figure~\ref{Fig:rotating_star_tracing} the most compact stable BS are in the right neighborhood of $\Omega_{max}$ if no ergoregion has appeared at that point. If $\Omega_{max}$ lies in the region where stars have ergoregions we consider the star where the ergoregion is just starting to appear (we shall call the scalar field frequency at this point $\Omega_{ergo}$), as we will see later the stars in the right neighborhood of $\Omega_{ergo}$ will be the most compact stable BS. We treat either $\Omega_{max}$ or $\Omega_{ergo}$ star as the most compact stable star keeping in mind that while the star at this exact point might technically be unstable, there exists a stable BS close by.
We note as we take lower values of $\sigma_0$ parameter we obtain more stiff shell-like solutions just as it happens in the spherically symmetric case \citep{Friedberg:1986tq}, although the shells now have toroid shapes instead of non-rotating spherical ones. Some of the star profiles are shown on the Figure~\ref{Fig:various_stars}. One important difference, however, is that unlike in spherically symmetric SBS case the compactness of the star at $\Omega_{max}$ doesn't increase as we decrease $\sigma_0$.

\subsection{Ergoregions}
\begin{figure}[H]
	\begin{center}
		\includegraphics[width=0.7\linewidth]{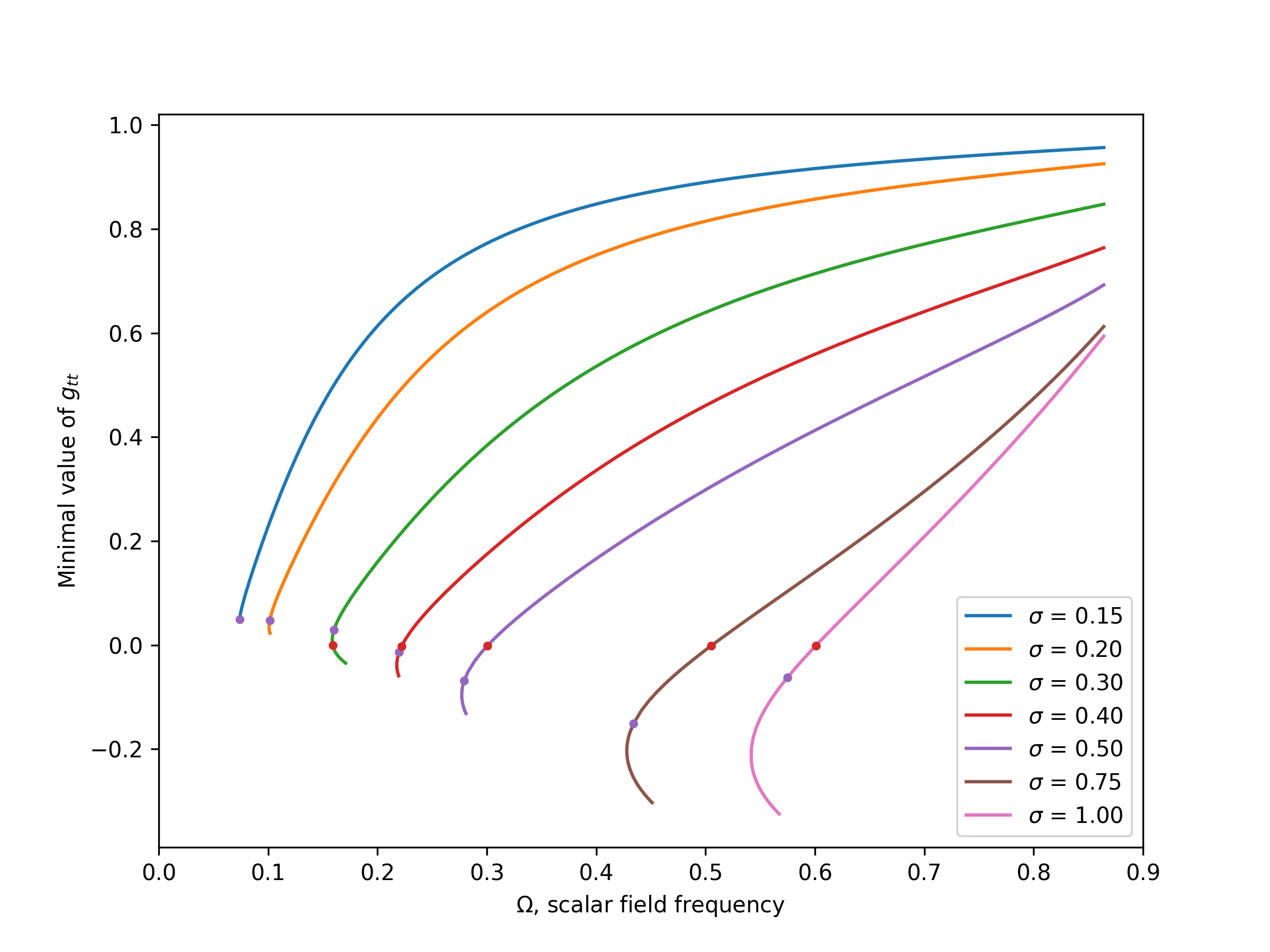}
	\end{center}
	\caption{The minimal value of $g_{tt}$ component of the SBS metric. The values of $g_{tt} < 0$ signify the presence of ergoregions.}
	\label{Fig:g_tt_component}
\end{figure}

\begin{figure}[H]
	\centering
	\subfloat{\includegraphics[width=0.5\linewidth]{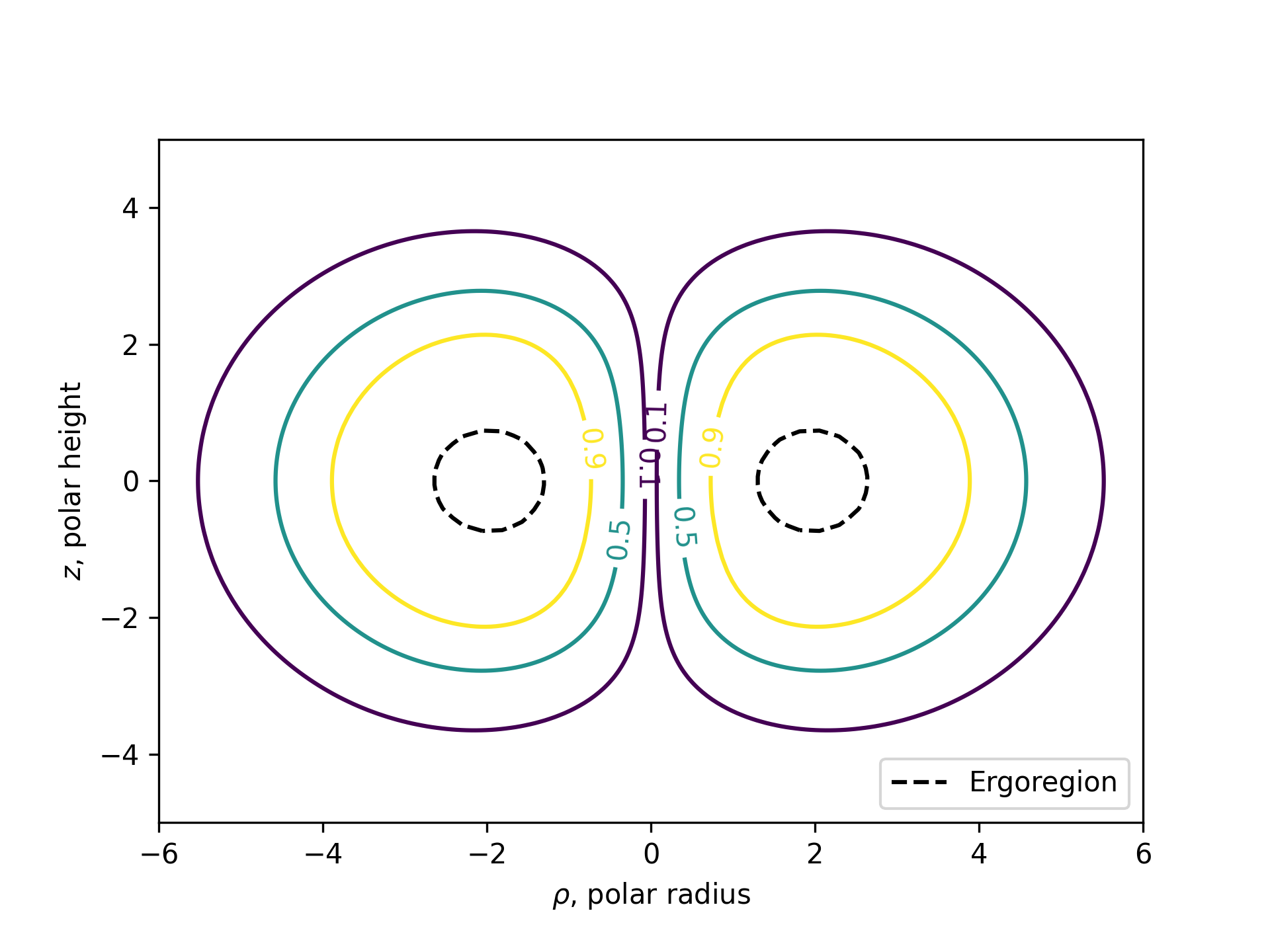}\centering}
	\hfill
	\subfloat{\includegraphics[width=0.5\linewidth]{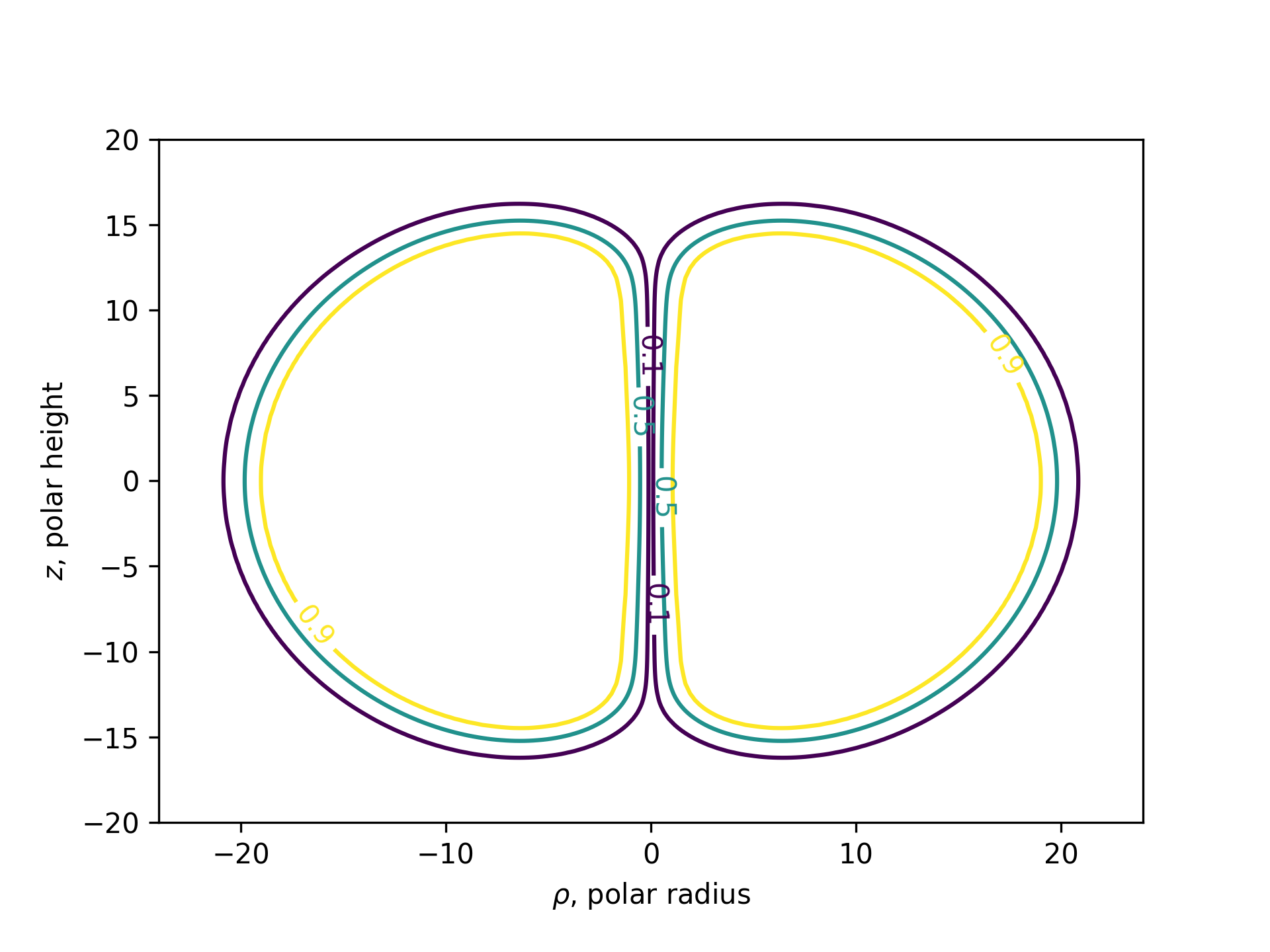}\centering}
	\caption{A less stiff ($\sigma_0 = 0.4$) star with an ergoregion and the more stiff ($\sigma_0 = 0.2$) star without an ergoregion. Both stars are taken at the global maximum mass point $\Omega_{max}$ of SBS family. The figures use cylindrical coordinates \eqref{Result:cylindrical_coordinates}. The level lines on the figures indicate the levels of constant absolute value of the scalar field, the values are given in fractions of $\phi_c$.}
	\label{Fig:ergoregion_stars}
\end{figure}

Another important feature of the rotating BS is the potential presence of ergoregions. The ergoregion of a compact object is the region where the spacetime spin (characterized by $g_{t\phi}$ metric component) is so high that all physical observers get dragged to rotate with the object, it occurs when the spacetime metric $g_{tt}$ component becomes negative. Typically ergoregions are bad news since they are known to be a source of instability \citep{Cardoso:2007az}. To detect the presence of ergoregions we examine the global minimum of $g_{tt}$ component of the metric, we demonstrate it on the Figure~\ref{Fig:g_tt_component}.
We see that the stars with ergoregions lie in the interval $\Omega < \Omega_{ergo}$, where $\Omega_{ergo}$ is the star with the zero minimal value of $g_{tt}$. As we move to the smaller values of $\sigma_0$ the region of SBS families with ergoregions moves closer to the global mass maximum. Finally, when $\sigma_0 = 0.3$ this region occurs beyond the maximal mass point in the section where SBS are certainly unstable ($\Omega_{ergo} < \Omega_{max}$). The examples of a star with ergoregion and without ergoregion is given on Figure~\ref{Fig:ergoregion_stars}.

\subsection{Multipole moments}
\begin{figure}[H]
	\begin{center}
		\includegraphics[width=1\linewidth]{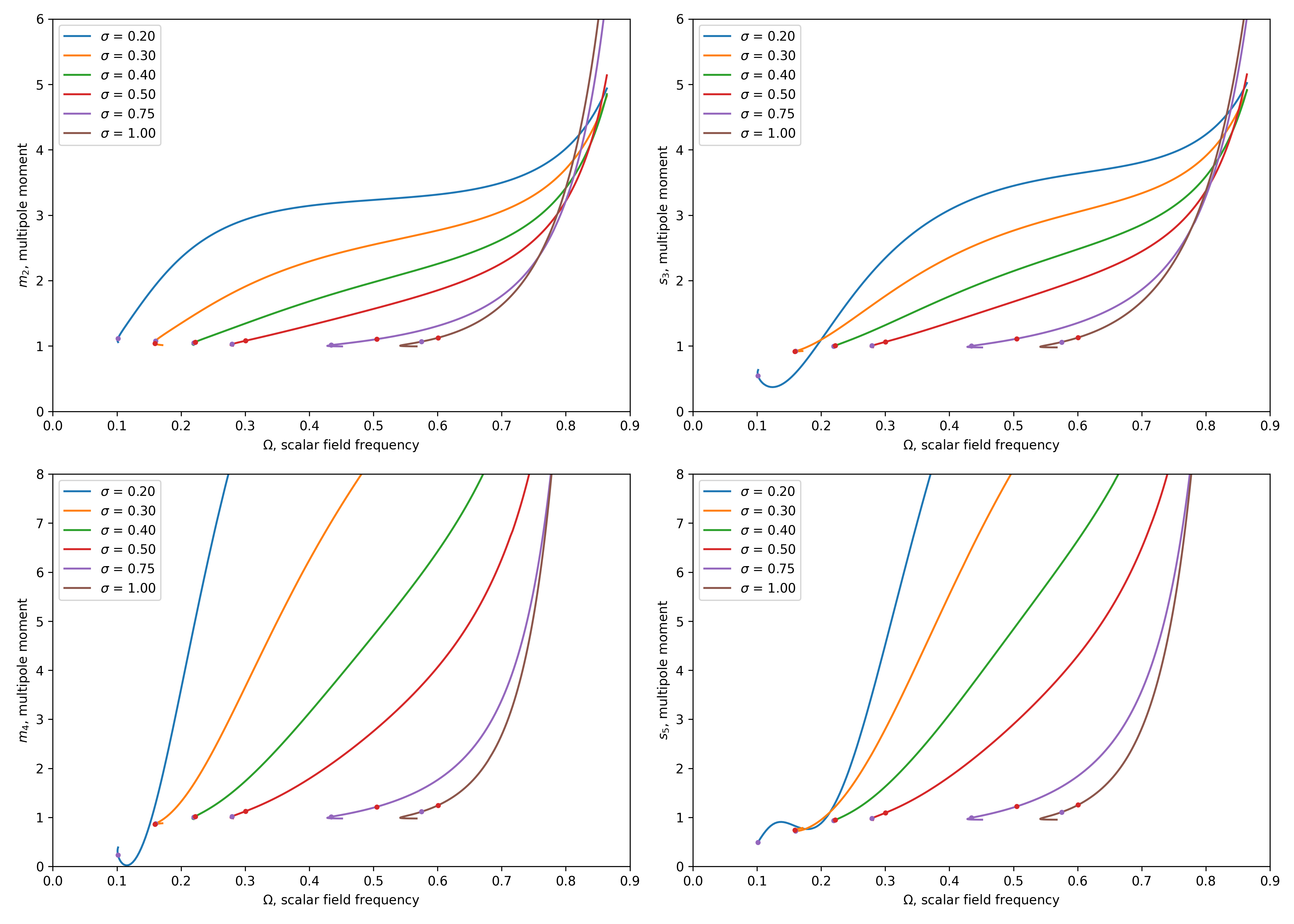}
	\end{center}
	\caption{The first four GH multipole moments $m_2$, $s_3$, $m_4$, $s_5$ for the star families with different values of $\sigma_0$. The multipole moments for the compact stars for the intermediate range $\sigma_0 \in [0.3, 0.5]$ are close to Kerr BH moments.}
	\label{Fig:multipole_moments}
\end{figure}

\begin{figure}[H]
	\begin{center}
		\includegraphics[width=1\linewidth]{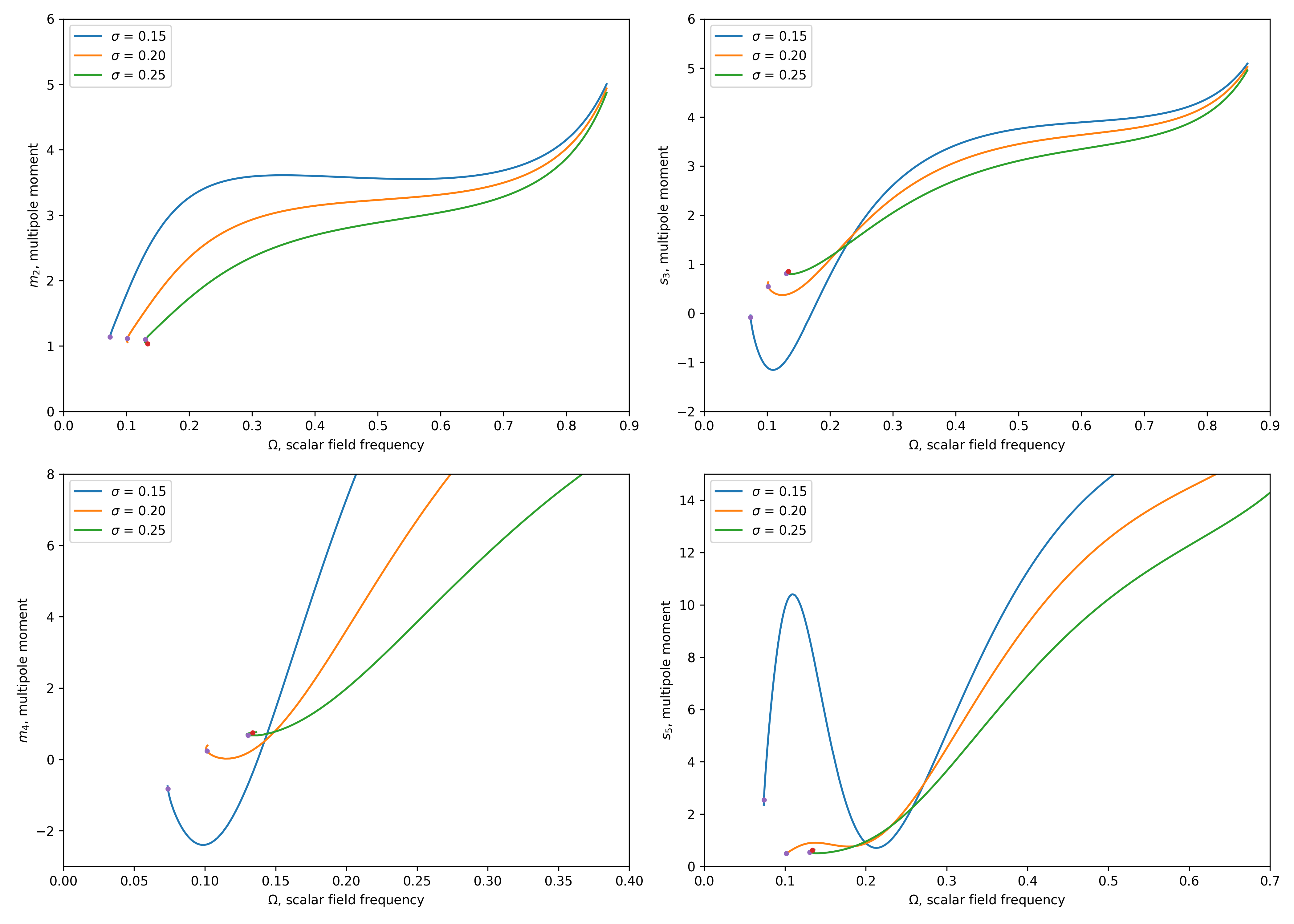}
	\end{center}
	\caption{The first four GH multipole moments $m_2$, $s_3$, $m_4$, $s_5$ for the star families with small parameter $\sigma_0 = \{0.15, 0.2, 0.25\}$. The multipole moments for the compact stars deviate from Kerr BH moments significantly.}
	\label{Fig:multipole_moments_extreme}
\end{figure}

To investigate how close SBS can resemble rotating BH we measure $m_2$, $s_3$, $m_4$ and $s_5$ reduced GH moments. As we discussed earlier, GH moments can be used as the measure of closeness of the spacetime outside the star to the rotating BH spacetime. The nonzero reduced GH moments of the Kerr metric are $m_{2n} = s_{2n+1} = 1$, where $n\in\mathbb{N}$. We demonstrate the reduced GH moments on the Figure~\ref{Fig:multipole_moments}.
We see that \textit{all} four investigated multipole moments for most compact stable stars located close to $\Omega_{max}$ in the intermediate range $\sigma_0 \in [0.3, 0.5]$ are very close to unity and therefore the spacetimes outside SBS are very close to Kerr spacetime (the GH moments of the stars with $\sigma_0 > 0.5$ also approach unity, but that section of SBS family is unstable). A surprising result, however, is that as $\sigma_0$ gets smaller and star profiles become stiffer, the multipole moments start to deviate significantly from Kerr values. This becomes especially evident on Figure~\eqref{Fig:multipole_moments_extreme}. We see that for $\sigma_0 = 0.15$ the moments $s_3$ and $m_4$ become negative long before $m_2$ approaches unity.

\section{Conclusions}
We surveyed a series of SBS families for a range of potential parameter $\sigma_0$ that correspond to the $m = 1$ azimuthal number branch. We found that as $\sigma_0$ approaches zero rotating SBS become more shell-like similar to their non-rotating counterparts, however unlike spherically symmetric SBS that take form of a spherical shell, the rotating SBS take form of a toroid shell.

While non-rotating SBS become more and more compact as $\sigma_0$ approaches zero, the stiffer rotating stars actually become less compact. SBS families with relatively high values of $\sigma_0$ have stars that contain ergoregions, however, we found that as $\sigma_0$ decreases the first occurrence of ergoregions is pushed closer and closer to the mass extremum, where SBS are expected to become unstable, and already at $\sigma_0 = 0.3$ all SBS with ergoregions are located in the unstable region.

We find that the stable SBS with the exterior spacetime most closely resembling rotating Kerr BH actually occur for the intermediate range $\sigma_0 \in [0.3, 0.5]$, where all the multipole moments for the most compact stable stars simultaneously approach Kerr values. However, for $\sigma_0 < 0.2$ the most compact stable stars significantly deviate from Kerr metric. It becomes progressively more numerically challenging to explore the SBS family branches when $\sigma_0 \leq 0.1$ while physically these stars are less interesting so we decided to stop the investigation at $\sigma_0 = 0.15$.

We can conclude that there exists a region of the parameter space with $\sigma_0 \in [0.3, 0.5]$ where the $m = 1$ branch SBS are simultaneously very compact $\mathcal{C} \approx 0.4$, don't have ergoregions, are likely stable and the spacetime metric outside these stars closely mimics Kerr metric. 

\section{Acknowledgments}
We are hugely indebted to Frans Pretorius for his patient guidance throughout this project. We also thank Justin Ripley, Alex Pandya, Alexey Milekhin and Oleksandr Stashko for useful discussions.

\section{Appendix}
\subsection{System of equations, boundary conditions and conserved charges}
We substitute the scalar field ansatz \eqref{Model:Scalar_field_ansatz} and the metric ansatz \eqref{Model:KEH_ansatz} into the Klein-Gordon equation \eqref{Model:KG_equations} and get the following equation
\begin{equation}
	\begin{split}
		&\phi_{,rr} + \left(\frac{2}{r}+\gamma_{,r}\right)\phi_{,r}+\frac{1}{r^2}\left(\phi_{,\theta\theta} + \left(\frac{\cos\theta}{\sin\theta}+\gamma_{,\theta}\right)\phi_{,\theta}\right)\\
		&+e^{2\sigma}\bigg[e^{-\gamma}\left(e^{-\rho}(\Omega + m\, \omega)^2 - e^\rho \frac{m^2}{r^2\sin^2\theta}\right)-V'\bigg]\phi = 0,\qquad V' = \frac{d V}{d |\Phi|^2}.
	\end{split}
\end{equation}
The Einstein equations \eqref{Model:Einstein_equations} together with the Klein-Gordon equation lead to a set of seven non-linear differential equations only five of which are independent. We take the following four equations together with the Klein-Gordon equation to form a system that fully describes a rotating Boson Star
\begin{subequations}
	\begin{align}
		&\gamma_{,rr}+\left(\frac{3}{r}+\gamma_{,r}\right)\gamma_{,r}+\frac{1}{r^2}\left[\gamma_{,\theta\theta}+\left(2\frac{\cos\theta}{\sin\theta}+\gamma_{,\theta}\right)\gamma_{,\theta}\right] - 2 \kappa e^{2\sigma}S_\gamma = 0,\\
		&\rho_{,rr} + \left(\frac{2}{r}+\gamma_{,r}\right)\rho_{,r}+\frac{1}{r^2}\left[\rho_{,\theta\theta}+\left(\frac{\cos\theta}{\sin\theta}+\gamma_{,\theta}\right)\rho_{,\theta}\right]-\frac{1}{r}\left(\gamma_{,r}+\frac{1}{r}\frac{\cos\theta}{\sin\theta}\gamma_{,\theta}\right)\\\nonumber
		&- e^{-2\rho}(r^2 \omega_{,r}^2 + \omega_{,\theta}^2)\sin^2\theta - 2 \kappa e^{2\sigma} S_\rho= 0,\\
		&\sigma_{,rr}+\frac{1}{r}\sigma_{,r} + \frac{1}{r^2}\sigma_{,\theta\theta} - \frac{1}{4}\left(\frac{2}{r}+\gamma_{,r}\right)\gamma_{,r}-\frac{1}{4 r^2}\left(2\frac{\cos\theta}{\sin\theta}+\gamma_{,\theta}\right)\gamma_{,\theta}\\\nonumber
		&-\frac{1}{4}\left(\frac{2}{r}-\rho_{,r}\right)\rho_{,r}-\frac{1}{4r^2}\left(2\frac{\cos\theta}{\sin\theta}-\rho_{,\theta}\right)\rho_{,\theta} -  \frac{1}{4}e^{-2\rho}(r^2 \omega_{,r}^2 + \omega_{,\theta}^2)\sin^2\theta\\\nonumber
		&+ \kappa e^{2\sigma} S_\sigma = 0,\\
		&\omega_{,rr} + \left(\frac{4}{r} + \gamma_{,r} - 2\rho_{,r}\right) \omega_{,r} + \frac{1}{r^2}\left[\omega_{,\theta\theta}+\left(3\frac{\cos\theta}{\sin\theta}+\gamma_{,\theta}-2\rho_{,\theta}\right)\omega_{,\theta}\right] \\\nonumber
		&- 2\kappa e^{2\sigma} S_\omega = 0,
	\end{align}
\end{subequations}
where the sources are
\begin{subequations}
	\begin{align}
		&S_\gamma = e^{-\gamma}\left[e^{-\rho}(\Omega + m\, \omega)^2 - e^{\rho}\frac{m^2}{r^2 \sin^2\theta}\right]\phi^2 - V,\\
		&S_\rho = e^{-\gamma}\left[e^{-\rho}(\Omega + m\, \omega)^2 + e^{\rho}\frac{m^2}{r^2 \sin^2\theta}\right]\phi^2,\\
		&S_\sigma = e^{-\gamma}\left[e^{-\rho}(\Omega + m\, \omega)^2 - e^{\rho}\frac{m^2}{r^2 \sin^2\theta}\right]\phi^2 + e^{-2\sigma}\left(\phi_{,r}^2 + \frac{1}{r^2}\phi_{,\theta}^2\right),\\
		&S_\omega = \frac{2}{r^2 \sin^2\theta}e^{\rho - \gamma} m (\Omega + m\,\omega)\phi^2,
	\end{align}
\end{subequations}
We use the remaining two equations as independent residuals to control the numerical error and numerical algorithm convergence
\begin{subequations}
	\begin{align}
		&\gamma_{,r\theta} + \frac{1}{2}\gamma_{,r}\gamma_{,\theta} + \left(\frac{1}{2}\frac{\cos\theta}{\sin\theta} - \sigma_{,\theta}\right)\gamma_{,r} - \left(\frac{1}{2r}+\sigma_{,r}\right)\gamma_{,\theta}-\frac{1}{2}\frac{\cos\theta}{\sin\theta}\rho_{,r}+\frac{1}{2}\rho_{,r}\rho_{,\theta}\\\nonumber
		&-\frac{1}{2r}\rho_{,\theta}  - \frac{\cos\theta}{\sin\theta}\sigma_{,r} - \frac{1}{r}\sigma_{,\theta} - \frac{1}{2}e^{-2\rho}\omega_{,r}\omega_{,\theta}\, r^2 \sin^2\theta + \kappa e^{2\sigma} \hat{S}^r_{\,\theta} = 0,\\
		&\gamma_{,rr}+\frac{1}{2}\left(\frac{2}{r}+\gamma_{,r}\right)\gamma_{,r}+\sigma_{,rr}-\gamma_{,r}\sigma_{,r}+\frac{1}{r^2}\left(\sigma_{,\theta\theta}+\left(\frac{\cos\theta}{\sin\theta}+\gamma_{,\theta}\right)\sigma_{,\theta}\right)\\\nonumber
		&-\frac{1}{2}\left(\frac{2}{r}-\rho_{,r}\right)\rho_{,r}-\frac{1}{2}e^{-2\rho}\omega_{,r}^2 r^2 \sin^2\theta + \kappa e^{2\sigma} \hat{S}^r_{\; r} = 0,
	\end{align}
\end{subequations}
where
\begin{subequations}
	\begin{align}
		&\hat{S}^r_{\; \theta} = 2 e^{-2\sigma} \phi_{,r}\phi_{,\theta},\\
		&\hat{S}^r_{\; r} = 2 e^{-2\sigma} \phi_{,r}^2 + V.
	\end{align}
\end{subequations}

The complete system of equations given above implies a set of regularity conditions for rotating stars $m\neq 0$. Furthermore, the Boson Star solution should be regular and free of angular defect at the axis of symmetry $\theta = 0$, which means that the spatial part of the metric \eqref{Model:KEH_ansatz} on the axis of symmetry must be conformal to the flat metric, which implies \cite{StephaniKramer:2003}
\begin{equation}
	\left[2\sigma + \gamma - \rho = 0\right]_{\theta = 0} = 0.
\end{equation}
Finally, we require the metric to be asymptotically flat and the scalar field to vanish at infinity. The reflection symmetry allows us to consider the solution on the halved interval $\theta \in [0, \pi/2]$. All these conditions can be put together in the following complete system of boundary conditions
\begin{itemize}
	\item At the origin $r = 0$
	\begin{equation}
		\left.\partial_r \gamma\right|_{r = 0} = \left.\partial_r \rho\right|_{r = 0} = \left.\partial_r \omega\right|_{r = 0} = 0,\qquad
		\left[\gamma - \rho - 2\sigma\right]_{r = 0} = 0,\qquad \left.\phi\right|_{r = 0} = 0.
	\end{equation}
	\item At the infinity $r \to \infty$
	\begin{equation}
		\left.\gamma\right|_{r\to \infty} = \left.\rho\right|_{r\to \infty} = \left.\sigma\right|_{r\to \infty} = \left.\omega\right|_{r\to \infty} = \left.\phi\right|_{r\to \infty} = 0.
	\end{equation}
	\item At the pole $\theta = 0$
	\begin{equation}
		\left.\partial_\theta \gamma\right|_{\theta = 0} = \left.\partial_\theta \rho\right|_{\theta = 0} = \left.\partial_\theta \omega\right|_{\theta = 0} = 0,\qquad
		\left[\gamma - \rho - 2\sigma\right]_{\theta = 0} = 0,\qquad \left.\phi\right|_{\theta = 0} = 0.
	\end{equation}
	\item At the equator $\theta = \pi/2$
	\begin{equation}
		\left.\partial_\theta \gamma\right|_{\theta = \frac{\pi}{2}} = \left.\partial_\theta \rho\right|_{\theta = \frac{\pi}{2}} = \left.\partial_\theta \sigma\right|_{\theta = \frac{\pi}{2}} = \left.\partial_\theta \omega\right|_{\theta = \frac{\pi}{2}} = \left.\partial_\theta \phi \right|_{\theta = \frac{\pi}{2}}= 0.
	\end{equation}
\end{itemize}

To find the expressions for the Komar mass and the angular momentum \eqref{Model:mass_momentum} we consider the hypersurface $t = const$. The normal vector to the hypersurface is given by
\begin{equation}
	n_\mu = -\{e^{\frac{1}{2}(\gamma + \rho)}, 0, 0, 0\},
\end{equation}
and the volume element is given by
\begin{equation}
	dV = e^{2\sigma + \frac{1}{2}(\gamma - \rho)}2\pi r^2 \sin\theta dr d\theta.
\end{equation}
Substituting the above expressions together with the ansatze (\ref{Model:Scalar_field_ansatz}, \ref{Model:KEH_ansatz}) we arrive at
\begin{subequations}
	\begin{align}
		&M = \frac{\kappa}{2}\int_\Sigma e^{2\sigma}\left(2e^{-\rho}\Omega(\Omega + m\omega)\phi^2 - e^{\gamma} V\right) r^2 \sin\theta dr d\theta,\\
		&J = - \frac{m \kappa}{2} \int_\Sigma e^{2\sigma-\rho }\left(\Omega + m \omega\right)\phi^2 r^2 \sin\theta dr d\theta.
	\end{align}
\end{subequations}

\subsection{Gerosh-Hansen moments}
We give a brief review of Gerosh-Hansen (GH) multipole moments \citep{Geroch:1970cd, Hansen:1974zz} for axisymmetric spacetimes and compute first few leading moments for a widely used metric \citep{Komatsu:1989zz}
\begin{equation}
	\label{KEH_metric}
	ds^2 = - e^{\gamma + \rho}dt^2 + e^{2\sigma}(dr^2 + r^2 d\theta^2) + e^{\gamma - \rho} r^2 \sin^2\theta (d\varphi - \omega dt)^2.
\end{equation}
We first give a brief overview of GH moments. We assume that the matter distribution that generates the metric \eqref{KEH_metric} is localized and either vanishes, or decays exponentially after a certain radius. The starting point for defining any multipole moments is to define a potential with respect to which the moments are then computed. In stationary spacetimes there exist two such potentials, $\phi_M$ and $\phi_J$, analogous to mass and angular momentum potentials in Newtonian theory \citep{Hansen:1974zz}
\begin{subequations}
	\begin{align}
		&\phi_M = \frac{1}{4\lambda}(\lambda^2 + \psi^2 - 1),\\
		&\phi_J = \frac{1}{2\lambda}\psi,
	\end{align}
\end{subequations}
where $\lambda$ is the norm and $\psi$ is the twist of the timelike Killing vector $\xi$. The potential $\phi_M$ reduces to Newtonian potential in the weak field limit, while $\phi_J$ lacks a similar straightforward interpretation. The choice of the potentials is not unique, however, for the given potentials the multipole moments of Kerr spacetime have an especially nice form. The norm is defined as
\begin{equation}
	\lambda = - \xi^a \xi_a.
\end{equation}
Next we consider the following expression
\begin{equation}
	\label{d_twist}
	\psi_a = \epsilon_{a b c d}\xi^b\triangledown^c\xi^d.
\end{equation}
Using the identity $\triangledown_a\triangledown_b \xi_c = \xi^d R_{d a b c}$ that follows from the Killing equation $\triangledown_a \xi_b + \triangledown_{b} \xi_a = 0$ we get
\begin{equation}
	\triangledown_{[a} \psi_{b]} = - \epsilon_{a b c d}\xi^c R^d_{\;e} \xi^e,
\end{equation}
where $R_{a b}$ is the Ricci tensor. In the vacuum spacetime case this makes $\psi_a$ curl-free and thus, at least locally, we can define the twist $\psi$ such that $\psi_a = \partial_a \psi$.

The timelike Killing vector $\xi^a$ for the metric \eqref{KEH_metric} is simply $\xi^a = \{1,0,0,0\}$, which allows us to easily find $\lambda$ explicitly as
\begin{equation}
	\lambda = e^\gamma\left(e^\rho - e^{-\rho}\omega^2 r^2 \sin^2\theta\right).
\end{equation}
However, since the definition of $\psi$ requires the metric to be Ricci flat, which is not necessarily true for a generic metric of the form \eqref{KEH_metric}, there is no explicit equation for $\psi$.

The next step in the construction is to establish a 3-space analogous to Euclidean space. Unfortunately, in the stationary spacetime case the timelike Killing vector field $\xi^a$ does not define a family of "moment of time" 3-surfaces since $\xi^a$ is not orthogonal to any surface family. However, if $M$ is a manifold with the metric $g_{ab}$ and a Killing field $\xi^a$ one can use the manifold $S$ of trajectories of the Killing vector field instead \citep{Gerosh:1971}. There exists a one-to-one correspondence between any tensor field $T'^{\;a\dots b}_{\;c \dots d}$ on the $S$ and a tensor field $T^{\;a\dots b}_{\;c \dots d}$ on $M$ provided that the latter satisfies
\begin{equation}
	\mathcal{L}_\xi T^{\;a\dots b}_{\;c \dots d} = 0,\quad \xi_a T^{\;a\dots b}_{\;c \dots d} = 0\quad \dots\quad \xi^d T^{\;a\dots b}_{\;c \dots d} = 0.
\end{equation}
The metric on $S$ corresponds to
\begin{equation}
	h_{ab} = \lambda g_{ab} + \xi_a \xi_b,
\end{equation}
for the metric \eqref{KEH_metric}, the metric on $S$ can be taken as
\begin{equation}
	\label{3D_S_metric}
	d\sigma^2 = e^{2\sigma + \gamma} \left(e^\rho - e^{-\rho}\omega^2 r^2 \sin^2\theta\right) (dr^2 + r^2 d\theta^2) + e^{2\gamma} r^2 \sin^2\theta d\phi^2.
\end{equation}

The existence of multipole moments in Newtonian theory depends in an essential way on the conformal flatness of Euclidean space, in General Relativity one can only expect to establish a multipole moment construction in the vicinity of the spatial infinity in an asymptotically flat spacetime \cite{Geroch:1970cd}. For that we first need to define what is asymptotic flatness. We are going to call the manifold $S$ asymptotically flat if we there is a larger manifold $\tilde{S}$ such as 
\begin{enumerate}
	\item $\tilde{S} = S \cup \Lambda$, where $\Lambda$ is a single point, corresponding to spatial infinity,
	\item $\tilde{h}_{ab} = \Omega^2 h_{ab}$ is a smooth metric on $\tilde{S}$,
	\item $\left.\Omega\right|_\Lambda = 0$,  $\left.\tilde{D}_a \Omega\right|_\Lambda = 0$, $\left.\tilde{D}_a\tilde{D}_b\Omega\right|_\Lambda = \left.2 \tilde{h}_{ab}\right|_\Lambda$, where $\tilde{D}_a$ is the covariant derivative associated with $\tilde{h}_{ab}$.
\end{enumerate}
The metric \eqref{3D_S_metric} of $S$ satisfies these conditions provided that the spacetime metric \eqref{KEH_metric} describes a localized object, it is sufficient to take $\Omega = 1/r^2$. It is then convenient to introduce a coordinate $\bar{r} = 1/r$ which is regular in the vicinity of and zero at $\Lambda$.

It is useful to rescale the potentials $\phi_M$ and $\phi_J$ as $\tilde{\phi}_{M,J} = \Omega^{-\frac{1}{2}}\phi_{M,J}$. When done in Newtonian theory such rescaling would essentially make the moments the derivatives of the potential in terms of coordinates regular at $\Lambda$ (e.g. a system with a single non-vanishing octupole moment $Q^{(3)}_{abc}$ in Cartesian coordinates $x^a$ would have the potential $\phi_N = Q^{(3)}_{abc} \bar{x}^a\bar{x}^b\bar{x}^c$, where $\bar{x}^a = x^a/{x^2}$ in accordance to our previous prescription for $\bar{r} = 1/r$).

The Newtonian multipole moments also obey certain translation relations. Knowing all multipole moments up to order $n$ using point $p$ as origin we can obtain all moments up to the same order $n$ at any other origin point $p'$. Translations under the inversion $\bar{x}^a = x^a/{x^2}$ become special conformal transformations, therefore the moments defined at the infinity $\Lambda$ should satisfy an appropriate set of transformation relations under conformal transformations. To make the multipole moments conform to these transformations in an asymptotically flat space, as shown in \citep{Geroch:1970cd}, one has to define the moments as 
\begin{equation}
	\label{Moments_recur}
	\begin{split}
		&P = \tilde{\psi},\\
		&P_{a_1\dots a_{s+1}} = \mathcal{T}\left[\tilde{D}_{a_1}P_{a_2\dots a_{s+1}} - \frac{1}{2}s(2s - 1) \tilde{\mathcal{R}}_{a_1 a_2} P_{a_3\dots a_{s+1}}\right],
	\end{split}
\end{equation}
where $\tilde{\mathcal{R}}_{a_1 a_2}$ is the Ricci tensor on $\tilde{S}$. In our case it is computed for the metric \eqref{3D_S_metric} after appropriate conformal transformation $\tilde{h}_{ab} = \Omega^2 h_{ab}$ in $(\bar{r},\theta,\phi)$ coordinates.

The spacetime \eqref{KEH_metric} in addition to timelike Killing vector $\xi$ also has a spacelike Killing vector $\eta^a = \{0,0,0,1\}$, all multipole moments have to be invariant with respect to $\eta$ rotations. The multipole moments can therefore be fully described by the set of numbers $M_n$ and $S_n$ 
\begin{equation}
	M_n = \frac{1}{n!} \left.M_{a_1\dots a_n} z^{a_1} \dots z^{a_n}\right|_\Lambda,\qquad S_n = \frac{1}{n!} \left.J_{a_1\dots a_n} z^{a_1} \dots z^{a_n}\right|_\Lambda,
\end{equation}
where $\left.z^a\right|_\Lambda = \{\cos\theta, - \frac{1}{x}\sin\theta, 0\}$ is a unit vector parallel to the axis of symmetry and $M_{a_1\dots a_n}$ and $J_{a_1\dots a_n}$ are corresponding moment tensors constructed from $\phi_M$ and $\phi_J$ respectively. Furthermore, due to reflection symmetry all odd $M_n$ moments and all even $S_n$ moments vanish.

Hansen \citep{Hansen:1974zz} shows that multipole moments of the Kerr spacetime are
\begin{equation}
	\label{Kerr_moments}
	M_{2n} = (-1)^{n + 1} M a^{2n},\qquad M_{2n+1} = 0,\qquad S_{2n} = 0,\qquad S_{2n+1} = (-1)^n M a^{2n + 1},
\end{equation}
where $M$ is the black hole mass and $a = J/M$ is the spin parameter. It is convenient to normalize the multipole moments using Kerr values and introduce the dimensionless reduced moments \cite{Yagi:2014bxa}
\begin{equation}
	m_{2n} = (-1)^{n+1} \frac{M_{2n}}{M^{2n+1} j^{2n}},\qquad s_{2n+1} = (-1)^n \frac{J_{2n+1}}{M^{2n+2} j^{2s+1}},
\end{equation}
where $j = J/M^2 = a/M$. These parameters are unit for Kerr.

Having defined the multipole moments we now demonstrate how to connect it to asymptotic expansions for the metric functions. Following general idea of \citep{Butterworth:1976,Komatsu:1989zz} we introduce a differential operator
\begin{equation}
	\triangle_{a} f = \nabla^2 f + \frac{a}{r}\partial_r f + \frac{a}{r^2}\frac{\cos\theta}{\sin\theta} \partial_\theta f = \partial_r^2 f + \frac{2 + a}{r} \partial_r f + \frac{1}{r^2}\left(\partial_\theta^2 f + (1 + a)\frac{\cos\theta}{\sin\theta}\partial_\theta f\right),
\end{equation}
where $\triangledown^2 f$ is a flat space Laplacian in spherical coordinates.
We can then express the equations for $\gamma$, $\rho$ and $\omega$ as
\begin{subequations}
	\label{KEH_method_equations}
	\begin{align}
		&\triangle_{0} \rho  =  2 \kappa e^{\gamma + 2\sigma}S_\gamma,\\
		\label{KEH_eq_exp_gamma}
		&\triangle_{1} (e^{\gamma}) = - \gamma_{,r} \rho_{,r} - \frac{1}{r^2}\gamma_{,\theta}\rho_{,\theta} + \frac{1}{r}\left(\gamma_{,r} + \frac{1}{r}\frac{\cos\theta}{\sin\theta}\gamma_{,\theta}\right) + e^{-2\rho}(r^2 \omega_{,r}^2 + \omega_{,\theta}^2)\sin^2\theta + 2 \kappa e^{2\sigma} S_\rho,\\
		&\triangle_{2} \omega = -\left(\gamma_{,r} - 2\rho_{,r}\right) \omega_{,r} - \frac{1}{r^2}\left(\gamma_{,\theta}-2\rho_{,\theta}\right)\omega_{,\theta} + 2\kappa e^{2\sigma} S_\omega.
	\end{align}
\end{subequations}
We are going to focus on the asymptotic region with negligible energy density or no energy density at all so we are assuming $S_\rho = S_\gamma = S_\omega = 0$. The angular eigenfunctions of the operator $\Delta_a$ can be expressed as Jacobi polynomials $P^{(\frac{a}{2},\frac{a}{2})}_n(\cos\theta)$, which inspires the following asymptotic metric decomposition
\begin{subequations}
	\label{KEH_asymptotics}
	\begin{align}
		&e^\gamma = 1 + \frac{B_0}{r^2} + \sum\limits_{n = 1}^\infty \frac{B_n}{r^{2n + 2}} P^{(\frac{1}{2},\frac{1}{2})}_n(\cos\theta),\\
		&\rho = \sum\limits_{n = 0}^\infty\left[\frac{\rho_n}{r^{2 n + 1}} + O(r^{-(2n + 2)}) \right] P_n(\cos\theta),\\
		&\omega = \sum\limits_{n = 0}^\infty\left[\frac{\omega_n}{r^{2n + 3}} + O(r^{-(2n + 4)}))\right] P^{(1,1)}_n(\cos\theta).
	\end{align}
\end{subequations}
Note that in $e^\gamma$ expression each Jacobi polynomial is multiplied by a single power of $r$ because the right hand side of the equation \eqref{KEH_eq_exp_gamma} vanishes, unfortunately this is not true for other metric functions. Applying Komar formulae to these expressions one gets $\rho_0 = - 2M$, $\omega_0 = 2 J$, where $M$ and $J$ are the mass and the angular momentum of the spacetime.

As in \citep{Butterworth:1976} we can substitute the asymptotic expansion \eqref{KEH_asymptotics} order by order into Einstein equations and find that the expansion uniquely depends on a family of parameters $B_n$, $\rho_n$ and $\omega_n$. E.g. to determine $M_2$ and $J_3$ moments it is sufficient to take
\begin{subequations}
	\begin{align}
		&e^{2\gamma} = 1 + \frac{B_0}{r^2} + \frac{B_2}{r^4}P^{(\frac{1}{2},\frac{1}{2})}_2(\cos\theta) + O\left(\frac{1}{r^6}\right),\\
		&\rho = - \frac{2 M}{r} - \frac{B_0}{r^2} + \frac{2 M B_0}{3 r^3} + \left(2 J^2 + \frac{B_0^2}{2} - \frac{5 B_2}{24}\right) \frac{1}{r^4} + \\\nonumber
		&\left[\frac{\rho_2}{r^3} - \left(4 J^2 + \frac{5 B_2}{3}\right) \frac{1}{r^4}\right] P_2(\cos\theta) + O\left(\frac{1}{r^5}\right),\\
		&\omega = \frac{2J}{r^3} - \frac{6 J M}{r^4} - \frac{6J}{5 r^5}(3 B_0 - 8 M^2) - J\left(\frac{32 M^3}{3} - \frac{40 B_0 M}{3} + \frac{2\rho_2}{5}\right)\frac{1}{r^6} + \\\nonumber
		&\bigg[\frac{\omega_2}{r^5} + \left(\frac{9 J \rho_2}{5} - \frac{5 M \omega_2}{2}\right)\frac{1}{r^6} -\bigg] P^{(1,1)}_2 (\cos\theta) + O\left(\frac{1}{r^7}\right),\\
		&\sigma = \frac{M}{r} - \left(\frac{M^2}{2} + B_0\right)\frac{1}{r^2} + \left(\frac{\rho_2}{4} - \frac{M B_0}{3}\right)\frac{1}{r^3} + \left(\frac{J^2}{4} - \frac{B_0^2}{2} + \frac{15 B_2}{8} - \frac{3 M\rho_2}{8}\right)\frac{1}{r^4} +\\\nonumber
		&\left[\left(\frac{M^2}{2} + 2 B_0\right)\frac{1}{r^2} - \frac{3\rho_2}{4r^3} + \left(-\frac{3 J^2}{2} + B_0 M^2 + 4 B_0^2 - 15 B_2 + \frac{9 M \rho_2}{4}\right)\frac{1}{r^4}\right]\cos^2\theta + O\left(\frac{1}{r^5}\right).
	\end{align}
\end{subequations}
These expressions applied to the definition of twist \eqref{d_twist} allow us to reconstruct asymptotic expressions for the $\phi_M$ and $\phi_J$ potentials
\begin{subequations}
	\begin{align}
		&\phi_M = - \frac{M}{r} - \left(\frac{\rho_2}{4} + \frac{2 M^3}{3} - \frac{M B_0}{3}\right)\frac{1}{r^3}  + \frac{3\rho_2}{4 r^3}\cos^2\theta + O\left(\frac{1}{r^5}\right),\\
		&\phi_J = \left[\frac{J}{r^2} - \left(\frac{9\omega_2}{8} - \frac{2 M^2 J}{5} + \frac{2 J B_0}{5}\right)\frac{1}{r^4}\right]\cos\theta + \frac{15\omega_2}{8 r^4} \cos^3\theta + O\left(\frac{1}{r^6}\right)
	\end{align}
\end{subequations}
Direct substitutions of the expansions above into the definition of the moments \eqref{Moments_recur} allows us to obtain the relations for multipole moments as functions of the expansion coefficients. We give the expressions for the first four nontrivial GH moments
\begin{subequations}
	\label{GH_moments}
	\begin{align}
		&M_2 = \frac{1}{2}\rho_2 + \frac{M^3}{3}\left(1 + 4 b_0\right),\\
		&J_3 = - \frac{3}{4}\omega_2 + \frac{3}{5} j M^4 \left(1 + 4 b_0\right),\\
		&M_4 = \frac{1}{2}\rho_4 - \frac{4}{7} \rho_2 M^2 \left(1 + 3 b_0\right) + M^5 \left[\frac{8}{7}b_2 - \frac{16}{5}b_0^2 - \frac{32}{31} b_0 + \frac{6}{35}j^2 - \frac{19}{105}\right],\\
		&J_5 = \frac{5}{6}\omega_4 - \frac{5}{6} \omega_2 M^2 \left(1 + 4 b_0\right) + \frac{5}{21}\rho_2 j M^3 - j M^6 \left[\frac{40}{21}b_2 - \frac{48}{7}b_0^2 - \frac{208}{63} b_0 + \frac{2}{7}j^2 - \frac{25}{63}\right],
	\end{align}
\end{subequations}
where $b_n = B_n/M^{2n+2}$ and $j = J/M^2$. 

The calculations and the resulting expressions are rather complex and in the past led to incorrect identification of the multipole moments \citep{Pappas:2012ns}. As a non-trivial check we take Kerr metric in Boyer-Lindquist coordinates, apply a coordinate transformation
\begin{equation}
	r = \hat{r} \left[\left(1 + \frac{M}{2\hat{r}}\right)^2 - \frac{a^2}{4 \hat{r}^2}\right],
\end{equation}
which is similar to the Schwarzschild isotropic coordinate transformation, to recast Kerr metric into \eqref{KEH_metric} form and identify Kerr asymptotic expansion parameters
\begin{equation}
	\begin{split}
		j = \frac{a}{M},\quad &b_0 = - \frac{1}{4} + \frac{a^2}{4 M^2},\quad \rho_2 = \frac{4}{3}M a^2,\quad \omega_2 = - \frac{8}{15}M a^3,\\
		&b_2 = 0,\quad \rho_4 = - \frac{16}{35} M a^4,\quad \omega_4 = \frac{16}{105} M a^5,
	\end{split}
\end{equation}
which, after substitution into \eqref{GH_moments}, leads to the correct expressions \eqref{Kerr_moments}.

There are two approaches to finding the coefficients $\rho_n$, $\omega_n$ and $B_n$ for a known numerical solution. As suggested by \citep{Vaglio:2022flq}, one could take the equations \eqref{KEH_method_equations}, obtain a formal solution to the equation $\Delta_a f = g$ (this is a part of a popular method used in fluid star calculations \citep{Komatsu:1989zz}) and then find intergral expression for the coefficients. The problem with this approach, however, is that the integral expressions have very high powers of $r$ and the resulting integrals are very hard to take numerically.

Alternatively, as suggested by \citep{Adam:2023qxj}, we can take the asymptotic expansions \eqref{KEH_asymptotics} and, using the orthogonality of Jacobi polynomials, take an angular integral of the corresponding metric functions with an appropriate factor to isolate the required mode. Then we take a radial limit numerically to separate $\rho_n$, $\omega_n$ or $B_n$ coefficients
\begin{subequations}
	\begin{align}
		&B_n = \frac{2n + 1}{2} \lim\limits_{r\to\infty} r^{n+2}\int\limits_0^\pi \left(e^{\gamma(r,\theta)}-1\right) P^{(\frac{1}{2},\frac{1}{2})}_n(\cos\theta) d\theta,\\
		&\rho_n = \frac{(n + 1)}{2} \frac{n! (n+1)!}{\Gamma\left(n + \frac{3}{2}\right)^2} \lim\limits_{r\to\infty} r^{n+1}\int\limits_0^\pi \rho(r,\theta) P_n(\cos\theta) d\theta,\\
		&\omega_n = \frac{(2n + 3)(n+2)}{8(n+1)} \lim\limits_{r\to\infty} r^{n+3}\int\limits_0^\pi \omega(r,\theta) P^{(1,1)}_n(\cos\theta) d\theta,
	\end{align}
\end{subequations}
where $P_n(\cos\theta) = P^{(0,0)}_n(\cos\theta)$ are Legendre polynomials.
We found this method to produce the most accurate numerical results.

\bibliographystyle{plain}
\bibliography{biblography}
\end{document}